\documentclass[11pt,a4paper]{iopart}
\usepackage{iopams}  
\usepackage{bm}
\begin{document}
\jl{14}
\newcommand{\be}{\begin{equation}}
\newcommand{\ee}{\end{equation}}
\newcommand{\ba}{\begin{eqnarray}}
\newcommand{\ea}{\end{eqnarray}}
\newcommand{\onehalf}{\textstyle{\frac{1}{2}}}
\def\qed{\hbox{${\vcenter{\vbox{
   \hrule height 0.4pt\hbox{\vrule width 0.4pt height 6pt
   \kern5pt\vrule width 0.4pt}\hrule height 0.4pt}}}$}}

\title[de Sitter special relativity]{de Sitter special relativity}

\author{R Aldrovandi, J P Beltr\'an Almeida and J G Pereira\footnote{jpereira@ift.unesp.br}}

\address{Instituto de F\'{\i}sica Te\'orica, 
Universidade Estadual Paulista, Rua Pamplona 145, 01405-900 S\~ao 
Paulo, Brazil}

\begin{abstract}
A special relativity based on the de Sitter group is introduced, which is the theory that might hold up in the presence of a non-vanishing cosmological constant. Like ordinary special relativity, it retains the quotient character of spacetime, and a notion of homogeneity. As a consequence, the underlying spacetime will be a de Sitter spacetime, whose associated kinematics will differ from that of ordinary special relativity. The corresponding modified notions of energy and momentum are obtained, and the exact relationship between them, which is invariant under a re-scaling of the involved quantities, explicitly exhibited. Since the de Sitter group can be considered a particular deformation of the Poincar\'e group, this theory turns out to be a specific kind of deformed (or doubly) special relativity. Some experimental consequences, as well as the causal structure of spacetime---modified by the presence of the de Sitter horizon---are briefly discussed.
\end{abstract}
\section{Introduction}

When the cosmological constant $\Lambda$ vanishes, absence of gravitation is represented by Minkowski spacetime, a solution of the sourceless Einstein's equation. Its isometry transformations are determined by the Poincar\'e group, which is also the group governing the kinematics of special relativity. For a non-vanishing $\Lambda$, however, Minkowski is no longer a solution of the corresponding Einstein's equation and becomes, in this sense, physically meaningless. In this case, absence of gravitation turns out to be represented by the de Sitter spacetime. Now, the group governing the kinematics in a de Sitter spacetime is not the Poincar\'e, but the de Sitter group. This means essentially that, for a non-vanishing $\Lambda$, ordinary Poincar\'e special relativity will no longer be valid, and must be replaced by a de Sitter special relativity. Since the local symmetry of spacetime will be represented by the de Sitter group, the tangent space at every point of spacetime must also be replaced by an osculating de Sitter space. This is the geometrical setting of a de Sitter special relativity. In this theory, due to the change in the kinematic group, the ordinary notions of energy and momentum, as well as the relationship between them, will change \cite{aap}. Furthermore, the causal structure of spacetime will also be modified by the presence of the de Sitter horizon. 

At the Planck scale, gravitation and quantum mechanics are somehow expected to meet. That scale is, in consequence, believed to be the threshold of a new physics. In particular, consistency arguments related to quantum gravity seem to indicate that Lorentz symmetry must be broken and ordinary special relativity might be no longer true \cite{hbar}. To comply with that violation without producing significant changes in special relativity far from that scale, the idea of a deformed (or doubly, as it has been called) special relativity (DSR) has been put forward  recently \cite{dsr}. In this kind of theory, Lorentz symmetry is deformed through the agency of a dimensional parameter $\kappa$, proportional to the Planck length.\footnote{For some reviews, as well as for the relevant literature, see Ref.~\cite{kg}.} Such a deformation implies that, in the high energy limit, a quantum theory of gravitation must be invariant, not under the Poincar\'e group, but under a ``$\kappa$-deformed'' Poincar\'e group which reduces to the standard Poincar\'e group in the low energy limit.

Now, the de Sitter group naturally involves a length parameter, which is related to the cosmological constant by Einstein's equations. In addition, since it has the Lorentz group as a subgroup, it can also be interpreted as a particular deformation of the Poincar\'e group. In fact, it is related to the Poincar\'e group through the contraction limit of a vanishing cosmological constant, in the very same way the Galilei group is related to the Poincar\'e group through the contraction limit of an infinite velocity of light. A special relativity based on the de Sitter group, therefore, gives rise to a kind of DSR.\footnote{Similar ideas have already been explored in Ref.~\cite{guoatall}.} The fundamental difference in relation to the usual DSR models is that, in this case, the equivalence between frames is ruled by the de Sitter group. As a consequence, the energy and momentum definitions will change, and will satisfy a generalized relation. Furthermore, since the Lorentz group is a sub-group of de Sitter, the Lorentz symmetry will remain as a sub-symmetry in the theory. The presence of the de Sitter length-parameter, however, in addition to modifying the symmetry group, modifies also the usual Lorentz causal structure of spacetime, defined by the light cone. In fact, the causal domain of any observer will be further restricted by the presence of an event horizon: the de Sitter horizon.

To get some insight on how a de Sitter special relativity might be thought of, let us briefly recall the relationship between the de Sitter and the Galilei groups, which comes from the Wigner--In\"on\"u processes of group contraction and expansion \cite{inonu,gil}. Ordinary Poincar\'e special relativity can be viewed as describing the implications to Galilei's relativity of introducing a fundamental velocity-scale into  the Galilei group. Conversely, the latter can be obtained from the special-relativistic Poincar\'e group by taking the formal limit of the velocity scale going to infinity (non-relativistic limit). We can, in an analogous way, say that de Sitter relativity describes the implications to Galilei's relativity of introducing both a velocity and a length scales in the Galilei group. In the formal limit of the length-scale going to infinity, the de Sitter groups contract to the Poincar\'e group, in which only the velocity scale is present. It is interesting to observe that the order of the group expansions (or contractions) is not important. If we introduce in the Galilei group a fundamental length parameter, we end up with the Newton-Hooke group \cite{nh}, which describes a (non-relativistic) relativity in the presence of a cosmological constant \cite{gibb}. Adding to this group a fundamental velocity scale, we end up again with the de Sitter group, whose underlying relativity involves both a velocity and a length scales. Conversely, the low-velocity limit of the de Sitter group yields the Newton-Hooke group, which contracts to the Galilei group in the limit of a vanishing cosmological constant.

A crucial property of the de Sitter relativity is that it retains the quotient character of spacetime and, consequently, a notion of homogeneity. As in special relativity, whose underlying  Minkowski spacetime is the quotient space of the Poincar\'e by the Lorentz groups, the underlying spacetime of the de Sitter relativity will be the quotient space of the de Sitter and the Lorentz groups. In other words, it will be a de Sitter spacetime. Now, a space is said to be transitive under a set of transformations --- or homogeneous under them --- when any two points of it can be attained from each other by a transformation belonging to the set. For example, the Minkowski spacetime is transitive under spacetime translations. The de Sitter spacetime, on the other hand, is found to be transitive under a combination of translations and proper conformal transformations, the relative importance of these contributions being determined by the value of the cosmological constant. We are here taking advantage of a common abuse of language, talking rather freely  of {\it the} de Sitter group, while allowing its length parameter to vary. Of course, what is meant is the family of all such groups, each one the group of motions of a de Sitter space with a different scalar curvature.

Due to its quotient character, spacetime will respond concomitantly to any deformation occurring in the symmetry group. For small values\footnote{The reference value for defining small and large $\Lambda$ is the Planck cosmological constant $\Lambda_P = 3/l_P^2$, with $l_P$ the Planck length. A small $\Lambda$ will then be characterized by $\Lambda \, l_P^2 \to 0$. A large $\Lambda$, on the other hand, will be characterized by $\Lambda \, l_P^2 \to 1$.} of $\Lambda$, for example, the underlying spacetime will approach Minkowski spacetime. In the contraction limit $\Lambda \to 0$, it is reduced to the flat Minkowski spacetime, which is transitive under ordinary translations. In the contraction limit $\Lambda \to \infty$, on the other hand, the underlying spacetime is reduced to  a new maximally-symmetric cone spacetime \cite{confor}, which is homogeneous under proper conformal transformations. It is important to remark that the $\Lambda \to \infty$ limit must be understood as purely formal. In fact, since a large $\Lambda$ means a small length parameter, quantum effects should necessarily be taken into account. Such effects, as is well known, provides a cut-off value for $\Lambda$, which prevents the limit to be physically achieved.

Motivated by the above arguments, the basic purpose of this paper is to develop a special relativity based on the de Sitter group. We will proceed as follows. Section 2 is a review of the fundamental properties of the de Sitter groups and spaces. Section 3 describes, for the sake of completeness, the main geometrical properties of the cone spacetime that emerges in the limit of an infinite cosmological constant. In section 4, the fundamentals of a de Sitter special relativity are presented and discussed. In particular, an analysis of the deformed group generators acting on the de Sitter space is made, which allows us to understand how a de Sitter relativity can give rise to an algebraically well defined theory. The modified notions of energy and momentum are obtained in section 5, and the new relationship between them explicitly exhibited. A discussion on the possible phenomenological implications is then presented. Finally, section 6 discusses the results obtained. 

\section{de Sitter spacetimes and groups}

\subsection{The de Sitter spacetimes}

Spacetimes with constant scalar curvature $R$ are maximally symmetric: they can lodge the highest possible number of Killing vectors. Given a metric signature, this spacetime is unique~\cite{weinberg} for each value of $R$. Minkowski spacetime $M$, with vanishing scalar curvature, is the simplest one. Its group of motion is the Poin\-ca\-r\'e group ${\mathcal P} = {\mathcal L} \oslash {\mathcal T}$, the semi-direct product of the Lorentz ${\mathcal L} = SO(3,1)$ and the translation group ${\mathcal T}$. The latter acts transitively on $M$ and its group manifold can be identified with $M$. Indeed,  Minkowski  spacetime is a homogeneous space under ${\mathcal P}$, actually the quotient
\[
M = {\mathcal P}/{\mathcal L}.
\]

Amongst curved spacetimes, the de Sitter and anti-de Sitter spaces are the only possibilities \cite{ellis}. One of them has negative, and the other has positive scalar curvature. They can be defined as hyper-surfaces in the ``host'' pseudo-Euclidean spaces ${\bf E}^{4,1}$ and ${\bf E}^{3,2}$, inclusions whose points in Cartesian coordinates $(\chi^A) = (\chi^0,
\chi^1, \chi^2, \chi^3, \chi^{4})$ satisfy, res\-pectively,
\[
\eta_{AB} \chi^A \chi^B \equiv (\chi^0)^2 - (\chi^1)^2 -
(\chi^2)^2 - (\chi^3)^2 - (\chi^{4})^2 = -\, l^2
\]
and
\[
\eta_{AB} \chi^A \chi^B \equiv (\chi^0)^2 - (\chi^1)^2 -
(\chi^2)^2 - (\chi^3)^2 + (\chi^{4})^2 = l^2,
\]
with $l$ the de Sitter length parameter. The Latin alphabet ($a, b, c \dots = 0,1,2,3$) will be used to denote the four-dimensional algebra and tangent space indices. Using then $\eta_{a b}$ for the Lorentz metric $\eta = $ diag $(1$, $-1$, $-1$, $-1)$, and the sign notation ${\sf s} = \eta_{44}$, the above conditions can be put together as
\be
\eta_{a b} \, \chi^{a} \chi^{b} + {\sf s} \, (\chi^4)^2 = {\sf s} \, l^2.
\label{dspace2}
\ee
Defining the dimensionless coordinate $\chi^{\prime 4} = \chi^4/l$, it becomes
\be
\frac{1}{l^2} \, \eta_{a b} \, \chi^{a} \chi^{b} + {\sf s} \, (\chi^{\prime 4})^2 = {\sf s}.
\label{dspace1}
\ee
For ${\sf s} = - 1$, we have the de Sitter space $dS(4,1)$, whose metric is induced from the pseudo-Euclidean metric $\eta_{AB}$ = $(+1,-1,-1,-1,-1)$. It has the pseudo-orthogonal group $SO(4,1)$ as group of motions. Sign ${\sf s} = + 1$ corresponds to  anti-de Sitter space, denoted by $dS(3,2)$. It comes from $\eta_{AB}$ = $(+1,-1,-1,-1,+1)$, and has $SO(3,2)$ as its group of motions. Both spaces are homogeneous~\cite{livro}:
\[
dS(4,1) = SO(4,1)/ {\mathcal L} \quad {\rm and} \quad dS(3,2) = SO(3,2)/ {\mathcal L}.
\]
In addition, each group manifold is a bundle with the corresponding de Sitter or anti-de Sitter space as base space and the Lorentz group ${\mathcal L}$ as fiber \cite{kono}. These spaces are solutions of the sourceless Einstein's equation, provided the cosmological constant $\Lambda$ and the length parameter $l$ are related by
\be
\Lambda = - \frac{3 {\sf s}}{l^2}.
\label{lambdaR}
\ee

\subsection{Stereographic coordinates}

For definiteness, as well as to comply with observational data \cite{obs}, we consider from now on the $SO(4,1)$  positive cosmological constant case. The de Sitter space is then defined by
\be
-\,  \frac{1}{l^2} \, \eta_{a b} \, \chi^{a} \chi^{b} + (\chi^{\prime 4})^2 = 1,
\label{dspace1b}
\ee
and the four-dimensional stereographic coordinates $\{x^a\}$ are obtained through a projection from the de Sitter hyper-surfaces into a target Minkowski spacetime. They are given by \cite{gursey}
\be
\chi^{a} = \Omega(x) \, x^a
\label{xix}
\ee
and
\be
\chi'^4 = -\, \Omega(x) \left(1 + \frac{\sigma^2}{4 l^2} \right),
\label{xi4}
\ee 
where
\be
\Omega(x) = \frac{1}{1 - {\sigma^2}/{4 l^2}} \, ,
\label{n}
\ee
with $\sigma^2 = \eta_{a b} \, x^a x^b$. The $\{x^a\}$ take values on the  Minkowski spacetime on which the stereographic projection is made.

\subsection{Kinematic groups: transitivity}

In terms of the host-space  Cartesian coordinates $\chi^A$, the generators of the infinitesimal de Sitter transformations are
\be
L_{A B} = \eta_{AC} \, \chi^C \, \frac{\partial}{\partial \chi^B} -
\eta_{BC} \, \chi^C \, \frac{\partial}{\partial \chi^A} \, \, .
\label{dsgene}
\ee
They satisfy the commutation relations
\be
\left[ L_{AB}, L_{CD} \right] = \eta_{BC} L_{AD} + \eta_{AD} L_{BC} - \eta_{BD} L_{AC}
- \eta_{AC} L_{BD}.
\label{desla}
\ee
In terms of the stereographic coordinates $\{x^a\}$, these generators are written as
\be
L_{ab} =
\eta_{ac} \, x^c \, P_b - \eta_{bc} \, x^c \, P_a
\label{cp0}
\ee
and
\be
L_{a4} = l P_a - ({4 l})^{-1} K_a,
\label{dstra}
\ee
where
\be
P_a = {\partial}/{\partial x^a}
\ee
are the translation generators (with dimension of {\it length}$^{-1}$), and
\be
K_a = \left(2 \eta_{ab} \, x^b x^c - \sigma^2 \delta_{a}{}^{c} \right) P_c
\ee
are the generators of {\it proper} conformal transformations (with dimension of {\it length}). 
Whereas $L_{a b}$ refer to the Lorentz subgroup of de Sitter, $L_{a 4}$ define transitivity on the corresponding de Sitter space. As implied by the generators (\ref{dstra}), the de Sitter spacetime is found to be transitive under a combination of translations and proper conformal transformations. The relative importance of each one of these transformations is determined by the value of the length parameter $l$  or, equivalently, by the value of the cosmological constant. It is important to remark once more that the generators $L_{a b}$ and $L_{a 4}$ provide a realization of the de Sitter transformations on Minkowski spacetime, the target space of the stereographic projection. Observe that the indices $a, b, c, \dots$ are raised and lowered with the Minkowski metric $\eta_{ab}$. 

\subsection{Contraction limits}

The art of group contraction involves the ability of placing beforehand the contraction parameters in appropriate positions. This is usually achieved by performing a similarity transformation in the original generators \cite{inonu2}. Furthermore, since these appropriate positions are different for different limits, the vanishing and the infinite limits of the cosmological constant must be considered separately as they require different parameterization.


\subsubsection{Vanishing cosmological constant limit}
To study the limit of a vanishing cosmological constant ($l \to \infty$), it is convenient to write the de Sitter generators in the form
\be
L_{ab} =
\eta_{ac} \, x^c \, P_b - \eta_{bc} \, x^c \, P_a
\label{dslore}
\ee
and
\be
\Pi_a \equiv \frac{L_{a 4}}{l} =
P_a - \frac{1}{4 l^2}\,  K_a.
\label{l0}
\ee
The generators $L_{ab}$ give rise to the usual Lorentz transformation in Minkowski spacetime, and satisfy the commutation relation
\be
\left[ L_{ab}, L_{cd} \right] = \eta_{bc} L_{ad} + \eta_{ad} L_{bc} - \eta_{bd} L_{ac}
- \eta_{ac} L_{bd}.
\ee
For $l \to \infty$, the generators $\Pi_a$ reduce to ordinary translations, and the de Sitter group contracts to the Poincar\'e group ${\mathcal P} = {\mathcal L} \oslash {\mathcal T}$. Concomitant with the algebra and group deformations, the de Sitter space $dS(4,1)$ reduces to the Minkowski spacetime
\[
M = {\mathcal P}/{\mathcal L},
\]
which is transitive under ordinary translations only.

\subsubsection{Infinite cosmological constant limit}

To begin with we recall that, in this limit, the de Sitter space tends to the conic spacetime, denoted $N$, which is related to Minkowski through the spacetime inversion \cite{confor}
\be
x^a \to -\,  \frac{x^a}{\sigma^2}.
\label{inversion}
\ee
In fact, under the spacetime inversion (\ref{inversion}), the points at infinity of $M$ are led to the vertex of the cone-space $N$, and those on the light-cone of $M$ become the infinity of $N$. Using this relation, by applying the duality transformation (\ref{inversion}) to the Minkowski interval
\be
ds^2 = \eta_{ab} \, dx^a dx^b,
\ee
we see that\footnote{In addition to denoting the indices of the Minkowski spacetime $M$, the Latin alphabet ($a, b, c \dots = 0,1,2,3$) will also be used to denote the algebra and space indices of the cone-spacetime $N$.}
\be
ds^2 \to d\bar{s}^2 = {\bar{\eta}}_{ab} \, d{x}^a d{x}^b,
\label{confin}
\ee
where
\be
{\bar{\eta}}_{ab} = {\sigma}^{-4} \, \eta_{ab}, \qquad {\bar{\eta}}^{ab} =
{\sigma}^{4} \, \eta^{ab}
\label{Nmetric}
\ee
is the metric on the cone-space $N$. It is important to recall also that the spacetime inversion (\ref{inversion}) is well known to relate translations with proper conformal transformations \cite{coleman}:
\be
P_a \rightarrow K_a.
\ee
The Lorentz generators, on the other hand, are found not to change:
\be
L_{ab} \to L_{ab}.
\ee

The above results imply that, to study the limit of an infinite cosmological constant ($l \to 0$), it is convenient to write the de Sitter generators in the form
\be
\bar{L}_{ab} \equiv \sigma^{-4} L_{ab} =
\bar{\eta}_{ac} \, x^c \, P_b - \bar{\eta}_{bc} \, x^c \, P_a
\label{dSLgbis}
\ee
and
\be
\bar{\Pi}_a \equiv 4 l \, L_{a 4} = 4 l^2 P_a - K_a.
\label{linf}
\ee
The generators $\bar{L}_{ab}$ satisfy the commutation relation
\be
\left[ \bar{L}_{ab}, \bar{L}_{cd} \right] = \bar{\eta}_{bc} \bar{L}_{ad} + \bar{\eta}_{ad} \bar{L}_{bc} - \bar{\eta}_{bd} \bar{L}_{ac} - \bar{\eta}_{ac} \bar{L}_{bd}.
\ee
Since the interval $d\bar{s}^2$  is conformally invariant, and since $\bar{L}_{ab}$ satisfy a Lorentz-like commutation relation, the latter can be interpreted as the generators of a {\it conformal Lorentz transformation}. For $l \to 0$, $\bar{\Pi}_a$ reduce to (minus) the proper conformal generators, and the de Sitter group contracts to the {\em conformal}\, Poincar\'e group $\bar{\mathcal P} = \bar{\mathcal L} \oslash \bar{\mathcal T}$, the semi-direct product of the conformal Lorentz $\bar{\mathcal L}$ and the {\it proper} conformal $\bar{\mathcal T}$ groups \cite{ap1}. Metric (\ref{Nmetric}) is actually invariant under ${\bar{\mathcal P}}$.
Concomitant with the above group contraction, the de Sitter spacetime reduces to the conic spacetime
\[
N = \bar{\mathcal P}/\bar{\mathcal L}.
\]
It is a new maximally symmetric spacetime, transitive under proper conformal transformations \cite{confor}.

\section{The cone-spacetime}

Before proceeding further, we present a glimpse of the general properties of the cone spacetime $N$. It represents an empty spacetime, in which all energy is in the form of dark energy \cite{cosmo1}. It is what a purely classical physics would lead to, and can be interpreted as the fundamental spacetime around which quantum fluctuations changing $l=0$ to $l=l_P$ (or equivalently, changing $\Lambda \sim \infty$ to $\Lambda = \Lambda_P =3/l_P^2$) would take place.

\subsection{Geometry}

The metric (\ref{Nmetric}) of the cone spacetime leads to the Christoffel components
\be
{\Gamma}^c{}_{ab} = 2 \sigma^{-2} x^d
(\eta_{ad} \, \delta^c{}_b + \eta_{bd} \, \delta^c{}_a \,
- \eta_{ab} \, \delta^c{}_d).
\label{lct}
\ee
In terms of $\bar{\eta}_{ab}$, it is written as
\be
{\Gamma}^c{}_{ab} \equiv
\bar{\Gamma}^c{}_{ab} = 2 \bar{\sigma}^{-2} x^d
(\bar{\eta}_{ad} \, \delta^c{}_b + \bar{\eta}_{bd} \, \delta^c{}_a \, - \bar{\eta}_{ab} \, \delta^c{}_d),
\label{lct2}
\ee
where $\bar{\sigma}^{2} = \bar{\eta}_{ab} \, x^a x^b$. As an easy calculation shows, the corresponding Riemann and Ricci curvatures vanish. In consequence, also the scalar curvature vanishes. Except at the origin, therefore, where the metric tensor is singular and the Riemann tensor cannot be defined, the cone $N$ is a flat spacetime.

\subsection{Killing vectors}

We are going now to solve the Killing equation for the conformal invariant metric $\bar{\eta}_{ab}$. The resulting vectors $\xi_a$ will be referred to as the {\it conformal Killing vectors}.\footnote{Not to be confused with the vectors solving the  conformal Killing equation $\mathcal{L}_{\xi} g_{\mu \nu} = \Omega^2 g_{\mu \nu}$.} The Killing equation $\mathcal{L}_{\xi} \bar{\eta}_{ab} = 0$, as usual, can be written in the form
\be
\bar{\nabla}_a \xi_b + \bar{\nabla}_b \xi_a = 0,
\ee
where $\bar{\nabla}_a$ is the covariant derivative in the connection $\bar{\Gamma}^c{}_{ab}$. Using Eq.~(\ref{lct2}), it can be rewritten as
\be
\bar{\eta}_{ac} \, \partial_b \xi^c + \bar{\eta}_{bc} \, \partial_a \xi^c + \bar{\eta}_{ab} \, \partial_c(\ln \bar{\sigma}^{-4}) \, \xi^c = 0.
\ee
The corresponding solution is
\be
\xi^a(x) = \alpha^c (\bar{\sigma}^2 \, \delta_c{}^a - 2 \bar{\eta}_{cd} \, x^d \, x^a ) + \beta^{ac} \, x_c,
\ee
with $\alpha^c$ and $\beta^{ac} = -\,  \beta^{ca}$ integration constants. We can then choose a set of ten Killing vectors as follows:
\be
\xi^a_{(c)}(x) = \bar{\sigma}^2 \, \delta_c{}^a - 2 \bar{\eta}_{cd} \, x^d \, x^a
\ee
and
\be
\xi^a_{(cd)}(x) = \delta^a{}_c \, x_d -  \delta^a{}_d \, x_c.
\ee
The four vectors $\xi^a_{(c)}(x)$ represent proper conformal transformations, whereas the six vectors $\xi^a_{(cd)}(x)$ represent spacetime rotations. The general Killing vector, therefore, is given by
\be
\xi^a(x) = \alpha^c \, \xi^a_{(c)}(x) + \beta^{cd} \, \xi^a_{(cd)}(x).
\ee
The existence of ten independent Killing vectors shows that the cone spacetime $N$ is, in fact, maximally symmetric.

\subsection{Casimir invariants}

Ordinary relativistic fields, and the particles which turn up as their quanta, are classified by representations of the Poincar\'e group ${\mathcal P} = {\mathcal L} \oslash {\mathcal T}$, which is of rank two. Each representation is, consequently,  fixed by the values of two Casimir invariants. As any functions of two invariants are also invariant, it is possible to choose two which have a clear relationship with simple physical characteristics: mass $(m)$ and spin $(s)$. Of all the families of representations of the Poincar\'e group \cite{wigner}, Nature seems to have given preference to one of the so-called discrete series, whose representations are fixed by the two invariants
\be 
C_2 = {\gamma}_{ab} \, P^a P^b = \qed \equiv -\, m^2 c^2
\ee
and
\be
C_4 = {\gamma}_{ab} \, {W}^{a} {W}^{b} \equiv -\, m^2 c^2 s (s+1),
\ee
with $W^a$ the Pauli-Lubanski vector
\be
{W}^a = \onehalf\,  {\epsilon}^{abcd} P_{b} S_{cd}. 
\ee
Any  metric ${\gamma}_{ab}$  invariant  under  the group action  would provide  invariants, but to arrive at the above mentioned physical choice, the Lorentzian metric ${\eta}_{ab}$ must be chosen. The first, involving only translation generators, fixes the mass. It defines the 4-dimensional Laplacian operator and, in particular, the Klein-Gordon equation
\be 
(\qed + m^2 c^2) \phi = 0,
\label{kg1}
\ee
which all relativistic fields satisfy. The second invariant is the square of the Pauli-Lubanski
operator, used to fix the spin. 

Analogously to the ordinary Poincar\'e group, the Casimir invariants of the conformal Poincar\'e group $\bar{\mathcal P} = \bar{\mathcal L} \oslash \bar{\mathcal T}$ can be constructed in terms of the metric ${\gamma}_{ab} = {\eta}_{ab}$, and of the generators ${S}_{ab}$ and ${K}_a$.\footnote{Alternatively, they can be obtained from the de Sitter Casimir invariants by taking the contraction limit $l \to 0$.} The first Casimir invariant is the norm of ${K}_a$,
\be
\bar{C}_2 = {\eta}_{ab} \, {K}^{a} {K}^{b} = \bar{\qed} = -\, \bar{m}^2 c^2,
\ee
where $\bar{m}$ is the conformal equivalent of the mass. If we identify $\partial^a \partial_a \equiv m^2$, we find that
\be
\bar{m}^2 = \sigma^4 \, m^2.
\ee
The conformal Klein-Gordon equation is consequently
\be
(\bar{\qed} + \bar{m}^2 c^2) \phi = 0.
\label{kg2}
\ee
The second Casimir invariant, on the other hand, is defined as
\be
\bar{C}_4 = {\eta}_{ab} \, \bar{W}^{a} \bar{W}^{b} = - \bar{m}^2 c^2 s (s + 1),
\ee
where $\bar{W}^a$ is the Pauli-Lubanski conformal-vector
\be
\bar{W}^a = \onehalf {\epsilon}^{abcd} {K}_{b} {S}_{cd}.
\ee

\section{The de Sitter special relativity}

We construct now a relativity theory based on the de Sitter group. In ordinary special relativity,  the underlying Minkowski spacetime appears as the quotient space between the Poincar\'e and the Lorentz groups. Similarly, in a de Sitter relativity, the underlying spacetime will be the quotient space between de Sitter and the Lorentz groups. This aspect is crucial, as it ensures the permanence of a notion of homogeneity. Instead of Minkowski space, however, the homogeneous spacetime will be, for positive $\Lambda$, the de Sitter spacetime $dS(4,1) = SO(4,1) / {\mathcal L}$.

The Greek alphabet ($\mu, \nu, \rho, \dots =0,1,2,3$) will be used to denote indices related to the de Sitter spacetime. For example, its coordinates will be denoted by $\{x^\mu\}$. We recall that the Latin alphabet ($a, b, c \dots = 0,1,2,3$) relate to the de Sitter algebra, as well as to the spacetime indices of both limits of the de Sitter spacetime: Minkowski, which appears in the limit of a vanishing $\Lambda$, and the cone spacetime, which appears in the limit of an infinite $\Lambda$. This allows the introduction of the hol\-o\-nomic tetrad $\delta^{a}{}_{\mu}$, which satisfies
\be
\eta_{\mu \nu} = \delta^{a}{}_{\mu} \delta^{b}{}_{\nu} \, \eta_{ab} \quad \mbox{and} \quad
\bar{\eta}_{\mu \nu} = \delta^{a}{}_{\mu} \delta^{b}{}_{\nu} \, \bar{\eta}_{ab}.
\ee
Consequently, we can also write
\be
\sigma^2 = \eta_{ab} \, x^a x^b = \eta_{\mu \nu} \, x^\mu x^\nu \quad \mbox{and} \quad
\bar{\sigma}^2 = \bar{\eta}_{ab} \, x^a x^b = \bar{\eta}_{\mu \nu} \, x^\mu x^\nu,
\ee
where we have identified $x^a = \delta^{a}{}_{\mu} x^\mu$. 

\subsection{Transitivity and the notion of distance}

The two concurrent, but different types of transformations appearing in the generators defining transitivity on the de Sitter spacetime give rise to two different notions of distance: one which is related to translations, and another which  is related to proper conformal transformations. This means that it is possible to define two different metrics in the de Sitter spacetime, one invariant under translations, and another invariant under proper conformal transformations. As a consequence, there will be two different family of geodesics, one joining all points equivalent under translations, and another joining all points equivalent under proper conformal transformations. If one considers only one of these families, therefore, there will be points in the de Sitter spacetime which cannot be joined to each other by any geodesic. This is a well known property of the de Sitter spacetime \cite{ellis}.

\subsubsection{Translational distance}

The first notion of distance is that related to translations. This notion will be important for small values of $\Lambda$, for which translations become the dominant part of the de Sitter transitivity generators. To study its properties, therefore, it is necessary to use a parameterization appropriate for the limit $\Lambda \to 0$.  This parameterization is naturally provided by Eq.~(\ref{dspace1}),
\be
K_G \, {\Omega}^2(x) \, {\sigma}^2 + (\chi'^4)^2 = 1,
\ee
where
\be
K_G = -\, 1/l^2
\ee
represents the Gaussian curvature of the de Sitter spacetime. We introduce now the an\-hol\-o\-no\-mic tetrad field
\be
h^a{}_\mu = \Omega \, \delta^a{}_\mu.
\label{tetraze}
\ee
If $\eta_{ab}$ denotes the Minkowski metric, the de Sitter metric can, in this case, be written as
\be
g_{\mu \nu}  \equiv   h^{a}{}_{\mu} \, h^{b}{}_{\nu} \, \eta_{a b} =
\Omega^2(x) \, \eta_{\mu \nu}.
\label{44}
\ee
It defines the ``translational distance'', with squared interval 
\be
d\tau^2 = g_{\mu \nu} \, dx^\mu dx^\nu \equiv
\Omega^2(x) \, \eta_{\mu \nu} \, dx^\mu dx^\nu.
\label{onod}
\ee
For $l \to \infty$ ($\Lambda \to 0$), it reduces to the Lorentz-invariant Minkowski interval
\be
d\tau^2 \to ds^2 = \eta_{\mu \nu} \, dx^\mu dx^\nu.
\ee

\subsubsection{Conformal distance}

The second notion of distance is that related to the proper conformal transformation. Since this transformation is the most important part of the transitivity generators for large values of $\Lambda$, its study requires a parameterization appropriate for the limit $\Lambda \to \infty$. This can be achieved by rewriting Eq.~(\ref{dspace1}) in the form
\be
\bar{K}_G\, \bar{\Omega}^2(x) \, \bar{\sigma}^{2} + (\chi'^4)^2 = 1,
\ee
where
\be
\bar{\Omega}(x) \equiv \frac{\sigma^2}{4 l^2} \, \Omega(x) = -\, 
\frac{1}{(1 - {4 l^2}/{\sigma^2} )}
\label{nbar}
\ee
is the new conformal factor, and
\be
\bar{K}_G = -\, 16 \, l^2
\ee
is the {\it conformal} Gaussian curvature. We introduce now the anholonomic tetrad field
\be
\bar{h}^a{}_\mu = \bar{\Omega}(x) \, \delta^a{}_\mu.
\label{tetrabar}
\ee
If $\bar{\eta}_{ab}$ denotes the cone spacetime metric, the corresponding de Sitter metric can, in this case, be written as
\be
\bar{g}_{\mu \nu} \equiv \bar{h}^a{}_\mu \bar{h}^b{}_\nu \, \bar{\eta}_{ab} =
\bar{\Omega}^2(x) \, \bar{\eta}_{\mu \nu}.
\label{confmet}
\ee
It defines the ``conformal distance'' on de Sitter spacetime, whose quadratic interval has the form
\be
d \bar{\tau}^2 \equiv \bar{g}_{\mu \nu} \, dx^\mu dx^\nu =
\bar{\Omega}^2(x) \, \bar{\eta}_{\mu \nu} \,  dx^\mu dx^\nu.
\ee
For $l \to 0$ ($\Lambda \to \infty$), de Sitter contracts to the cone spacetime $N$, and $d \bar{\tau}^2$ reduces to the conformal invariant interval on $N$:
\be
d \bar{\tau}^2 \to d\bar{s}^2 = \bar{\eta}_{\mu \nu} \, dx^\mu dx^\nu.
\ee
On account of the conformal transitivity of this spacetime, this is the only notion of distance that can be defined on $N$.

\subsubsection{Two family of geodesics}

The Christoffel connection of the de Sitter spacetime metric $g_{\mu \nu}$ is
\be
\Gamma^{\lambda}{}_{\mu \nu} = \left[ \delta^{\lambda}{}_{\mu}
\delta^{\sigma}{}_{\nu} + \delta^{\lambda}{}_{\nu}
\delta^{\sigma}{}_{\mu} - \eta_{\mu \nu} \eta^{\lambda \sigma} \right]
\partial_\sigma \left[\ln \Omega(x)\right].
\label{46}
\ee
The corresponding Riemann tensor is
\be
R^{\mu}{}_{\nu \rho \sigma} = -\, \frac{1}{l^2} \,
\left[\delta^{\mu}{}_{\rho} g_{\nu \sigma} - \delta^{\mu}{}_{\sigma} g_{\nu
\rho} \right].
\label{47}
\ee
If we consider the family of geodesics defined by the Christoffel connection (\ref{46}), there will be points in the de Sitter spacetime which cannot be connected by anyone of these geodesics. The reason for this fact is that the metric $g_{\mu \nu}$ defines a ``translational distance'' only, whereas the de Sitter spacetime is homogeneous under a combination of translation and proper conformal transformations.

On the other hand, the Christoffel connection of the de Sitter spacetime $\bar{g}_{\mu \nu}$ is
\be
\bar{\Gamma}^{\lambda}{}_{\mu \nu} = \left[ \delta^{\lambda}{}_{\mu}
\delta^{\sigma}{}_{\nu} + \delta^{\lambda}{}_{\nu}
\delta^{\sigma}{}_{\mu} - \bar{\eta}_{\mu \nu} \bar{\eta}^{\lambda \sigma} \right]
\partial_\sigma \left[\ln \bar{\Omega}(x)\right].
\label{46b}
\ee
Similarly, the corresponding Riemann tensor is
\be
\bar{R}^{\mu}{}_{\nu \rho \sigma} = - 16 l^2 \,
\left[\delta^{\mu}{}_{\rho} \bar{g}_{\nu \sigma} -
\delta^{\mu}{}_{\sigma} \bar{g}_{\nu \rho} \right].
\label{47b}
\ee
Since the metric $\bar{g}_{\mu \nu}$ defines only a ``conformal distance'', and since the de Sitter spacetime is homogeneous under a combination of translation and proper conformal transformations, there will again be points in the de Sitter spacetime which cannot be connected by anyone of the geodesics belonging to the family of the Christoffel connection (\ref{46b}). However, the two families of geodesics are complementary in the sense that the points that cannot be connected by one family of geodesics can be connected by the other family.

It is important to remark that both Riemann tensors $R^{\mu}{}_{\nu \rho \sigma}$ and $\bar{R}^{\mu}{}_{\nu \rho \sigma}$ represent the curvature of the de Sitter spacetime. The difference is that, whereas $R^{\mu}{}_{\nu \rho \sigma}$ represents the curvature tensor in a parameterization appropriate for studying the limit of a vanishing cosmological constant, $\bar{R}^{\mu}{}_{\nu \rho \sigma}$ represents the curvature tensor in a parameterization appropriate for studying the limit of an infinite cosmological constant. As a straightforward calculation shows, both limits yield a spacetime with vanishing curvature. This means that Minkowski and the cone spacetimes are both flat.

\subsection{The de Sitter transformations}

The de Sitter transformations can be thought of as rotations in a five-dimensional pseudo-Euclidian spacetime,
\be
{\chi'}^C = \Lambda^C{}_D \, \chi^D,
\ee
with $\Lambda^C{}_D$ the group element in the vector representation. Since these transformations leave invariant the quadratic form
\be
- \eta_{AB} \chi^A \chi^B = l^2,
\ee
they also leave invariant the length parameter $l$. Their infinitesimal form is
\be
\delta {\chi}^C = \onehalf \, {\mathcal E}^{AB}  L_{AB} \, \chi ^C,
\ee
where ${\mathcal E}^{AB}$ are the parameters and $L_{AB}$ the generators.

\subsubsection{Small cosmological constant}
\label{scc}

For $\Lambda$ small, analogously to the identifications (\ref{dslore}) and (\ref{l0}), we define the parameters
\be
\epsilon^{ab} = {\mathcal E}^{ab} \quad \mbox{and} \quad
\epsilon^a = l \, {\mathcal E}^{a4}.
\ee
In this case, in terms of the stereographic coordinates, the infinitesimal de Sitter transformation assumes the form
\be
\delta x^c = \onehalf \, \epsilon^{ab} L_{ab} x^c + \epsilon^a \Pi_a x^c,
\ee
or equivalently
\be
\delta x^c = \epsilon^{c}{}_a x^a + \epsilon^a - \frac{\epsilon^b}{4 l^2}
\left(2 x_b x^c - \sigma^2 \delta_b{}^c \right).
\ee
In the limit of a vanishing $\Lambda$, it reduces to the ordinary Poincar\'e transformation.

\subsubsection{Large cosmological constant}
\label{lcc}

For $\Lambda$ large, analogously to the identifications (\ref{dSLgbis}) and (\ref{linf}), we define the parameters
\be
\bar{\epsilon}^{ab} = \sigma^{4} \, {\mathcal E}^{ab} \quad \mbox{and} \quad
\bar{\epsilon}^{a} = {\mathcal E}^{a4} / 4 l.
\ee
In this case, in terms of the stereographic coordinates, the de Sitter transformation assumes the form
\be
\delta x^c = \onehalf \, \bar{\epsilon}^{ab} \bar{L}_{ab} \, x^c + \bar{\epsilon}^a \bar{\Pi}_a \, x^c,
\ee
or equivalently
\be
\delta x^c = \bar{\epsilon}^{c}{}_a x^a - \bar{\epsilon}^b
\left(2 x_b x^c - \sigma^2 \delta_b{}^c \right) + 4 l^2 \bar{\epsilon}^a,
\ee
where $ \bar{\epsilon}^{c}{}_a = \bar{\epsilon}^{cb} \, \bar{\eta}_{ba} \equiv {\epsilon}^{c}{}_a$. In the limit of an infinite $\Lambda$, it reduces to the a conformal Poincar\'e transformation.

\subsection{The Lorentz generators}

Up to now, we have studied the de Sitter transformations in a Minkowski spacetime. In what follows we are going to study the form of the corresponding generators in a de Sitter spacetime, which is the spacetime of a de Sitter special relativity. This will be done by contracting the generators acting in Minkowski spacetime with the appropriate tetrads. We begin by considering the Lorentz generators.

\subsubsection{Small cosmological constant}

For small $\Lambda$, the generators of an infinitesimal Lorentz transformation are (see section \ref{scc})
\be
L_{ab} = \eta_{ac} x^c P_b - \eta_{bc} x^c P_a.
\label{dSLg}
\ee
The corresponding generators acting on a de Sitter spacetime can be obtained by contracting $L_{ab}$ with the tetrad $h^a{}_\mu$, given by Eq.~(\ref{tetraze}):
\be
{\mathcal L}_{\mu \nu} \equiv h^a{}_\mu \, h^b{}_\nu \, L_{ab} =
g_{\mu \rho} \, x^\rho \, P_\nu - g_{\nu \rho} \, x^\rho \, P_\mu.
\label{dsgene2}
\ee
The corresponding matrix vector representation is easily found to be
\be
({\mathcal S}_{\mu \nu})_\lambda{}^\rho = g_{\mu \lambda} \, \delta_\nu{}^\rho -
g_{\nu \lambda} \, \delta_\mu{}^\rho.
\ee
The spinor representation, on the other hand, is
\be
({\mathcal S}_{\mu \nu}) = 
{\frac{i}{4}} [\gamma_\mu, \gamma_\nu],
\ee
where $\gamma_\mu = h^a{}_\mu \, \gamma_a$ are the point-dependent Dirac matrices. For $l \to \infty$, the de Sitter spacetime reduces to Minkowski, and the corresponding Lorentz generators reduce to the generators of the usual, Minkowski  spacetime Lorentz transformation.

Now, the generators ${\mathcal L}_{\mu \nu}$ satisfy the commutation relation
\be
\left[ {\mathcal L}_{\mu \nu}, {\mathcal L}_{\rho \lambda} \right] = g_{\nu \rho} {\mathcal L}_{\mu \lambda} + g_{\mu \lambda} {\mathcal L}_{\nu \rho} - g_{\nu \lambda} {\mathcal L}_{\mu \rho} - g_{\mu \rho} {\mathcal L}_{\nu \lambda}.
\ee
Even when acting on de Sitter spacetime, therefore, these generators still present a well-defined algebraic structure, isomorphic to the usual Lie algebra of the Lorentz group. This is a fundamental property in the sense that it allows the construction, on the de Sitter spacetime, of an algebraically well defined special relativity. This possibility is related to the mentioned fact that, like the Minkowski spacetime, the (conformally-flat) de Sitter spacetime is homogeneous and isotropic \cite{jack}.

\subsubsection{Large cosmological constant}

For $\Lambda$ large, the generators of infinitesimal Lorentz transformations are (see section \ref{lcc})
\be
\bar{L}_{ab} = \bar{\eta}_{ac} x^c P_b - \bar{\eta}_{bc} x^c P_a.
\label{dSLgbis2}
\ee
On a de Sitter spacetime, their explicit form can be obtained by contracting (\ref{dSLgbis2}) with the tetrad $\bar{h}^a{}_\mu$, given by Eq.~(\ref{tetrabar}):
\be
\bar{\mathcal L}_{\mu \nu} \equiv \bar{h}^a{}_\mu \, \bar{h}^b{}_\nu \, \bar{L}_{ab} =
\bar{g}_{\mu \rho} \, x^\rho \, P_\nu - \bar{g}_{\nu \rho} \, x^\rho \, P_\mu
\label{dsgene4}
\ee
These generators are easily found to satisfy the commutation relation
\be
\left[ \bar{\mathcal L}_{\mu \nu}, \bar{\mathcal L}_{\rho \lambda} \right] = \bar{g}_{\nu \rho} \bar{\mathcal L}_{\mu \lambda} + \bar{g}_{\mu \lambda} \bar{\mathcal L}_{\nu \rho} - \bar{g}_{\nu \lambda} \bar{\mathcal L}_{\mu \rho} - \bar{g}_{\mu \rho} \bar{\mathcal L}_{\nu \lambda}.
\ee
Like ${\mathcal L}_{\mu \nu}$, therefore, they present a Lorentz-like algebraic structure. The corresponding matrix vector representation is, in this case, given by
\be
(\bar{\mathcal S}_{\mu \nu})_\lambda{}^\rho = \bar{g}_{\mu \lambda} \, \delta_\nu{}^\rho -
\bar{g}_{\nu \lambda} \, \delta_\mu{}^\rho,
\label{eq:defmatrix}
\ee
whereas the spinor representation is
\be
(\bar{\mathcal S}_{\mu \nu}) =
{\frac{i}{4}} [\bar{\gamma}_\mu, \bar{\gamma}_\nu],
\label{eq:defspinor}
\ee
with $\bar{\gamma}_\mu = \bar{h}^a{}_\mu \, \gamma_a$ the point-dependent Dirac matrices. For $l \to 0$, the de Sitter spacetime reduces to the conic space $N$, and the corresponding Lorentz generators reduce to the generators of a conformal Lorentz transformation.

\subsection{The de Sitter ``translation'' generators}

Like in the case of the Lorentz generators, the form of the generators $\Pi^a$ and $\bar{\Pi}^a$ acting in the de Sitter spacetime can be obtained through contractions with the appropriate tetrad. For $\Lambda$ small, they are given by
\be
\Pi_\mu \equiv h^a{}_\mu \, \Pi^a = \Omega \left[P_\mu - (1/4 l^2)^{-1} K_\mu \right],
\ee
where
\be
P_\mu = \partial/\partial x^\mu \quad \mbox{and} \quad K_\mu = \left(
2 \eta_{\mu \rho} \, x^\rho x^\nu - \sigma^2 \delta_\mu{}^\nu \right) P_\nu.
\ee
For $\Lambda$ large, on the other hand, they are  
\be
\bar{\Pi}_\mu \equiv \bar{h}^a{}_\mu \, \bar{\Pi}^a = \bar{\Omega} \left(P_\mu - (1/4 l^2)^{-1} K_\mu \right).
\ee
We see from these expressions that the de Sitter spacetime is transitive under a combination of of the translation and proper conformal generators. For $\Lambda \to 0$, $\Pi_\mu$ reduce to the usual translation generators of Minkowski spacetime. For $\Lambda \to \infty$, $\bar{\Pi}_\mu$ reduce to the proper conformal generators, which define the transitivity on the cone spacetime.

\subsection{Understanding the de Sitter relativity}

The de Sitter special relativity can be viewed as made up of two different relativities: the usual one, related to translations, and a conformal one, related to proper conformal transformations. It is a single relativity interpolating these two extreme limiting cases. For small values of $\Lambda$, for example, usual special relativity will prevail over the conformal one, and the Poincar\'e symmetry will be weakly deformed. In the contraction limit of a vanishing cosmological constant, de Sitter relativity reduces to usual special relativity. The underlying spacetime in this case reduces to the Minkowski space $M$, which is transitive under translations only. For large values of $\Lambda$, on the other hand, conformal relativity will prevail over the usual one, and the Poincar\'e symmetry will be strongly deformed. In the contraction limit of an infinite $\Lambda$, de Sitter special relativity reduces to conformal relativity. The underlying spacetime, in this case, will be the cone-space $N$, which is transitive under proper conformal transformations only.

\subsubsection{Conformal relativity}

Conformal relativity is, therefore, the limit of de Sitter special relativity for an infinite cosmological constant. It is the special relativity governing the equivalence of frames in the cone spacetime $N$. Notice that this equivalence must be understood in the conformal sense. In fact, remember that two points of this spacetime cannot be related by a translation, but only by a proper conformal transformation. Accordingly, kinematics will be governed by the so called conformal Lorentz group, whose generators are
\be
\bar{L}_{ab} = \bar{\eta}_{ac} \, x^c \, P_a -
\bar{\eta}_{bc} \, x^c \, P_a.
\ee
The corresponding conformal vector and spinor matrix repre\-sen\-ta\-tions are the limiting cases of (\ref{eq:defmatrix}) and (\ref{eq:defspinor}),
\be
(\bar{S}_{ab})_d{}^c = \bar{\eta}_{ad} \, \delta_b{}^c -
\bar{\eta}_{bd} \, \delta_a{}^c
\ee
and
\be
\bar{S}_{ab} =
{\frac{i}{4}} \, [\bar{\gamma}_a, \bar{\gamma}_b],
\ee
where $\bar{\gamma}_a = -\, \sigma^{-2} \, \gamma_a$ is a kind of conformal Dirac matrix. Observe that the anti-commutator of the $\bar{\gamma}_a$'s yields the cone spacetime metric:
\be
\{\bar{\gamma}_a, \bar{\gamma}_b \} = 2 \, \bar{\eta}_{ab}.
\ee
Of course, like the cone spacetime $N$, this limiting theory has to be interpreted as purely formal. It is what a classical physics would lead to, that is to say, it is the classical relativity behind the quantum physics at the Planck scale.

\section{Energy-momentum relations}

\subsection{Noether current}

Let us consider now the mechanics of point particles on de Sitter spacetime. The conserved Noether current associated to a particle of mass $m$ is, in this case, the five-dimensional angular momentum \cite{gursey}
\be
\lambda^{A B} = m c \left(\chi^A \; \frac{d \chi^B}{d \tau} - \chi^B \; \frac{d \chi^A}{d \tau} \right),
\ee
with $d\tau$ the de Sitter line element (\ref{onod}). In order to make contact with the usual definitions of energy and momentum, we rewrite it in terms of the stereographic coordinates $\{x^a\}$ and the Minkowski interval $ds$. The result is 
\be
\lambda^{ab} = x^a \, p^b - x^b \, p^a
\label{cc0}
\ee
and
\be
\lambda^{a4} = l p^a - (4 l)^{-1} \, k^a,
\label{cc2}
\ee
where
\be
p^a = m \, c \, \Omega \, \frac{dx^a}{ds} 
\ee
is the momentum, and
\be
k^a = (2 \eta_{cb} \, x^c \, x^a -
\sigma^2 \, \delta_b{}^a) \,  {p}^b
\ee
is the so called conformal momentum \cite{coleman}. Their form on the de Sitter spacetime can be obtained through a contraction with appropriate tetrads.

\subsection{Small cosmological constant limit}

For $\Lambda \, l_P^2 \to 0$, analogously to the generators, we define the de Sitter momentum
\be
\pi^a \equiv \frac{\lambda^{a 4}}{l} =
{p}^a - \frac{{k}^a}{4 l^2}.
\label{dstra8}
\ee
The corresponding spacetime version is
\be
\pi^\mu \equiv h_a{}^\mu \, \pi^a =
{p}^\mu - \frac{{k}^\mu}{4 l^2},
\label{dstra4}
\ee
where
\be
p^\mu = m \, c \, \frac{dx^\mu}{ds} 
\ee
is the Poincar\'e momentum, and 
\be
k^\mu = (2 \eta_{\lambda \rho} \, x^\rho \, x^\mu -
\sigma^2 \, \delta_\lambda{}^\mu) \,  {p}^\lambda
\ee
is the corresponding conformal Poincar\'e momentum.\footnote{Analogously to the identification $p^\mu = T^{\mu 0}$, with $T^{\mu \nu}$ the energy-momentum current, the conformal momentum $k^\mu$ is defined by $k^\mu = K^{\mu 0}$, with $K^{\mu \nu}$ the conformal current \cite{coleman}.} We remark that $\pi^\mu$ is the conserved Noether momentum related to the transformations generated by $\Pi_a$. Its zero component,
\be
\pi^0 \equiv {p}^0 - \frac{{k}^0}{4 l^2},
\ee
represents the energy, whereas the space components $(i, j, \dots = 1, 2, 3)$
\be
\pi^i \equiv {p}^i - \frac{{k}^i}{4 l^2}
\ee
represent the momentum. The presence of a cosmological constant, therefore, changes the usual definitions of energy and momentum \cite{aap}. As a consequence, the energy-momentum relation will also be changed \cite{hossen}.

In fact, the energy-momentum relation in de Sitter relativity is given by
\be
g_{\mu \nu} \pi^\mu \pi^\nu = \Omega^2 \, \eta_{\mu \nu} \left(
p^\mu p^\nu - \frac{1}{2 l^2} p^\mu k^\nu + \frac{1}{16 l^4} k^\mu k^\nu \right).
\label{emr1}
\ee
The components of the Poincar\'e momentum $p^\mu$ are
\be
p^\mu = \left(\frac{\varepsilon_p}{c} , p^i \right),
\ee
where $\varepsilon_p$ and $p^i$ are the usual Poincar\'e energy and momentum, respectively. As is well known, they satisfy the relation $\eta_{\mu \nu} \, p^\mu p^\nu = m^2 c^2$, where $m^2 c^2$ is the first Casimir invariant of the Poincar\'e group. Analogously, the components of the conformal momentum $k^\mu$ can be written in the form
\be
k^\mu = \left(\frac{\varepsilon_k}{c} , k^i \right),
\ee
with $\varepsilon_k$ the conformal notion of energy, and $k^i$ the space components of the conformal momentum. The conformal momentum satisfies $\eta_{\mu \nu} \, k^\mu k^\nu = \bar{m}^2 c^2$, where $\bar{m}^2 c^2$ is the first Casimir invariant of the conformal Poincar\'e group. Using the expressions above, the relation (\ref{emr1}) becomes
\be
\frac{\varepsilon_p^2}{c^2} - {p}^{2} = m^2  c^2 + \frac{1}{2 l^2} \left[
\frac{\varepsilon_p \varepsilon_k}{c^2} - \vec{p} \cdot \vec{k} - m \bar{m} c^2 -
\frac{1}{8 l^2} \left(\frac{\varepsilon_k^2}{c^2} - k^2 - \bar{m}^2 c^2 \right) \right].
\label{hedr}
\ee
For small values of $\Lambda$, the de Sitter length parameter $l$ is large, and the modifications in the energy-momentum relation will be small. In the limit of a vanishing cosmological constant, the ordinary notions of energy and momentum are recovered, and the de Sitter relativity reduces to the ordinary special relativity, in which the Poincar\'e symmetry is exact. The energy-momentum relation, in this case, reduces to the usual expression
\be
\frac{\varepsilon_p^2}{c^2} - {p}^2 = m^2 \, c^2.
\label{poincaredr}
\ee

\subsection{High cosmological constant limit}

For $\Lambda \, l_P^2 \to 1$, analogously to the generators, we define the de Sitter momentum
\be
\pi^a \equiv 4 l \, {\lambda^{a 4}} =
4 l^2 {p}^a - {k}^a.
\label{dstra9}
\ee
The corresponding spacetime version is
\be
\bar{\pi}^\mu \equiv \bar{h}_a{}^\mu \, \bar{\pi}^a =
\frac{4 l^2}{\sigma^2} \left( 4 l^2 {p}^\mu - {k}^\mu \right).
\label{dstra10}
\ee
We remark that $\bar{\pi}^\mu$ is the conserved Noether momentum related to the transformations generated by $\bar{\Pi}_a$. Its zero component,
\be
\bar{\pi}^0 = \frac{4 l^2}{\sigma^2} \, ( 4 l^2 {p}^0 - {k}^0),
\ee
represents the conformal energy, whereas the space components
\be
\bar{\pi}^i = \frac{4 l^2}{\sigma^2} \, ( 4 l^2 {p}^i - {k}^i)
\ee
represent the conformal momentum.

The energy-momentum relation is now given by
\be
\bar{g}_{\mu \nu} \bar{\pi}^\mu \bar{\pi}^\nu = 16 l^4 \, \bar{\Omega}^2 \, \sigma^{-8} \, \eta_{\mu \nu} \left[16 l^4 p^\mu p^\nu - 8 l^2 p^\mu k^\nu + k^\mu k^\nu \right].
\label{emr5}
\ee
In terms of the energy and momentum components, it becomes
\be
\frac{\varepsilon_k^2}{c^2} - {k}^2 = \bar{m}^2 c^2 + 8 l^2 \left[
\frac{\varepsilon_p \varepsilon_k}{c^2} - \vec{p} \cdot \vec{k} - m \bar{m} c^2 -
2 l^2 \left(\frac{\varepsilon_p^2}{c^2} - p^2 - m^2 c^2 \right) \right].
\label{ledr}
\ee
For large values of the cosmological constant, the de Sitter length parameter $l$ is small. 
In the formal limit $\Lambda \, l_P^2 \to \infty$, only the conformal notions of energy and momentum will remain, and de Sitter relativity will reduce to the pure conformal special relativity. In this case, the energy-momentum relation will be
\be
\frac{\varepsilon_k^2}{c^2} - {k}^2 = \bar{m}^2 \, c^2.
\ee

\subsection{Some conceptual issues}

In the standard formulations of deformed special relativity, which are based on a $\kappa$-deformed Poincar\'e group, the momentum space of the particles is identified with a de Sitter space \cite{kgn}. In these theories, although energy and momentum keep their special relativistic notions, they satisfy a deformed dispersion relation. As a consequence, a consistent notion of total energy and momentum, as well as a conservation law for them, is lacking \cite{kgn}. In the de Sitter special relativity, on the other hand, a precise notion of momentum and energy is provided for the particles: they are the Noether currents associated to de Sitter symmetry. This means that there is a clear relation between the symmetry generators (\ref{cp0}-\ref{dstra}) and the conserved currents (\ref{cc0}-\ref{cc2}). The resulting deformed dispersion relations for the particle's energy and momentum, given by Eqs.~(\ref{hedr}) and (\ref{ledr}), are consequently relations between conserved quantities. Since the de Sitter current is a linear combination of the momentum $p^\mu$ and the conformal current $k^\mu$, the dispersion relation turns out to depend on these two four-vectors. Notice, however, that neither $p^\mu$ nor $k^\mu$ is conserved: only the de Sitter momentum is conserved.

Another relevant feature of the de Sitter modified dispersion relations concerns their properties under a re-scaling of the fundamental quantities. In the very same way as it happens with the ordinary special relativity dispersion relation (\ref{poincaredr}), the de Sitter dispersion relations (\ref{hedr}) and (\ref{ledr}) are invariant under a simultaneous re-scaling of mass, energy and momentum. On the other hand, because it includes non-quadratic terms in the momentum, the dispersion relations of the usual formulations of DSR are not invariant under such a re-scaling. This invariance is an important issue because it prevents the so called ``soccer-ball problem'' \cite{soccer}. This means that the dispersion relation of the de Sitter DSR is true for elementary particles (as it is argued since they must be relevant for discussing elementary particle processes), as well as for macroscopic objects, like for example a soccer-ball. 

\subsection{Experimental consequences and speculations}

The change from Minkowski to a de Sitter spacetime implies a change in the symmetry group of spacetime from Poincar\'e to de Sitter. Since, algebraically speaking, the only difference between these groups is the replacement of $P_a$ by a certain combination of $P_a$ and $K_a$, the net result of this change is ultimately the {\it breakdown of translational invariance}. From the experimental point of view, therefore, a de Sitter special relativity may be probed by looking for possible violations of translational invariance in high energy processes. This can be done by applying the same procedure used in the search for possible violations of Lorentz and CPT symmetries in high energy processes \cite{lcpt}. For small values of $\Lambda$, as we have seen, the homogeneity of spacetime is preponderantly given by the translation generators, which means that the violation of the translation invariance will be very small. Only when $\Lambda$ is large this violation is expected to be relevant.

Now, relying on our current theories of particle physics based on spontaneously broken symmetry and phase transitions, there must have been some periods in the history of the universe in which the value of $\Lambda$, and hence of the scale energy $E_{\Lambda}$, were large. For example, in the electroweak epoch characterized by $\Lambda_{EW}$, the kinematics of a typical electroweak process with energy $E_{\Lambda_{EW}}$, according to the de Sitter special relativity, must have been strongly influenced by $\Lambda_{EW}$. In fact, as we have seen, a large cosmological constant would produce significant changes in the definitions of energy and momentum, as well as in the kinematic relations satisfied by them. These changes could modify significantly the physics that should be applied in the study the early universe.

On the other hand, if we take the phase transitions associated to the spontaneously broken symmetries as the primary source of a non-vanishing $\Lambda$, it is conceivable that a high energy experiment could modify the {\it local} structure of space-time for a {\it short period of time}, in such a way that the immediate neighborhood of a high energy collision would depart from the Minkowski space and become a de Sitter spacetime. According to this point of view, there would be a connection between the energy scale of the experiment and the local value of $\Lambda$ \cite{mansouri}.  It is interesting to note that such a connection yields a thermodynamical cutoff for the cosmological constant. To see it, observe first that the area of the de Sitter horizon is $A_{dS} \sim l^2$. Since the entropy associated to this surface is proportional to the logarithm of the number of states
\[
n = A_{dS} / l_P^2 \sim {l^2}/{l_P^2},
\]
with $l_P$ the Planck length, and since the minimum allowed value for the entropy is achieved for $n=1$, we see that the minimum allowed value for $l$ is of the order of the Planck length. This relation provides a contact between de Sitter special relativity and quantum gravity \cite{nzcc}.

Assuming the above described connection between the energy scale of the ex\-per\-iment and the local value of $\Lambda$, it is possible to envisage some potential experimental consequences of the dispersion relations (\ref{hedr}) and (\ref{ledr}). Of course, at the cosmological level, where \cite{carrol}
\[
l_{\Lambda_{0}} \sim \Lambda_{0}^{-{1}/{2}} \sim
10^{28}~\mbox{cm},
\]
which corresponds to $E_{\Lambda_0} \sim 10^{-33}$~{eV}, the deviation from the dispersion relations are very small, and there is no hope for any experimental detection in the existing colliders. However, for energies of the order of 200~GeV, corresponding to the electroweak phase transition, the de Sitter parameter is $l_{\Lambda_{EW}} \sim ({1}/{4})$ cm, which is equivalent to $E_{\Lambda_{EW}}\sim 10^{-4}$~eV. For high energy experiments of order 20~TeV, one finds $E_{\Lambda_{TeV}} \sim 1$~eV. And for energies of order 1000~TeV, we have $E_{\Lambda}\sim 2500$~eV. For particles of small mass, such as neutrinos, there would be significant changes in the kinematics at very high energies, which could eventually be tested in a foreseeable future \cite{mansouri}.

\section{Final remarks}

If the cosmological constant $\Lambda$ has a non-vanishing value, {\it ordinary special relativity breaks down and must necessarily be replaced by a de Sitter special relativity}. A crucial point of this theory is that it preserves the notion of spacetime ho\-mog\-e\-neity. In fact, like Minkowski, the de Sitter spacetime is a quotient space: $dS(4,1) = SO(4,1) / {\mathcal L}$. As a consequence, any deformation occurring in the symmetry group will produce concomitant deformations in the quotient space. In particular, different values of the cosmological constant will give rise to different spacetimes. For small $\Lambda$, the de Sitter group approaches the Poincar\'e group, and the de Sitter spacetime will approach the Minkowski spacetime. For large $\Lambda$, on the other hand, the de Sitter group approaches the conformal Poincar\'e group, and the underlying spacetime will approach a flat cone space. If we consider a possible $\Lambda$-dependence of high energy processes \cite{mansouri} close to the Planck scale, not only the symmetry group will change, but also the geometric nature of spacetime will change. Transitivity properties, in special, will be completely different. Accordingly, the energy and momentum definitions will change, and will satisfy a generalized relation. Experimentally, these changes would appear as a violation of the translational symmetry, and could eventually be tested through the application of the same techniques already used in the search for possible violations of the Lorentz and CPT symmetries.

Another important point is that, due to the homogeneous character of the de Sitter spacetime, the Lorentz generators in this spacetime still present a well defined algebraic structure, isomorphic to the usual Lie algebra of the Lorentz group. This means that the Lorentz symmetry remains a sub-symmetry in a de Sitter relativity, and consequently the velocity of light $c$ is left unchanged by a de Sitter transformation. Since it also leaves unchanged the length parameter $l$, a de Sitter transformation leaves unchanged both $c$ and $l$. This property has important consequences for causality. As is well known, the constancy of $c$ introduces a causal structure in spacetime, defined by the light cones. Analogously, the presence of the de Sitter length parameter $l$ adds to that structure some further restrictions on the causal structure of spacetime. To see that, let us remember that the de Sitter spacetime has a horizon, which restricts the causal region of each observer. In terms of the stereographic coordinates, this horizon is identified by
\be
x^2 + y^2 + z^2 = l^2/\Omega^2 \quad \mbox{and}
\quad (x^0)^2 = l^2 (2 - 1/\Omega)^2.
\ee
For small $\Lambda$, the horizon tends to infinity, and there are no significant causal changes. For large values of $\Lambda$, however, the causal domain of each observer --- restricted by the horizon --- becomes small. Considering again a possible $\Lambda$-dependence of high energy processes, at the Planck scale this region would be of the order of the Planck length. At this scale, therefore, the large value of $\Lambda$ would introduce deep changes in the causal structure of spacetime. This mechanism could eventually be an explanation for the causal modifications expected to occur at the Planck scale.

Finally, it is worth mentioning a topic of special importance, which concerns relativistic fields. If relativity changes, the concept of relativistic field must change accordingly. For example, in the context of the de Sitter relativity, a scalar field should be interpreted as a singlet representation, not of the Lorentz, but of the de Sitter group. Among other consequences, the Klein-Gordon equation will have a different form. For general values of $\Lambda$, it is \cite{aap}
\be 
\qed\, \phi + m^2 c^2 \, \phi - \frac{R}{6}\, \phi = 0,
\label{ciKGe}
\ee
with $\qed$ the Laplace-Beltrami operator in the de Sitter metric (\ref{44}), and $R= - 12/l^2$. Notice in passing that this could be the solution to the famous controversy on the $R/6$ factor~\cite{AANSM97}. In fact, this factor appears naturally if, instead of a Lorentz scalar, field $\phi$ is assumed to be a de Sitter scalar, in which case the corresponding Klein-Gordon equation becomes naturally conformal invariant in the limit of a vanishing mass. The ordinary Klein-Gordon equation (\ref{kg1}) for a Lorentz scalar is recovered in the limit $\Lambda \to 0$. For large values of $\Lambda$, on the other hand, the Klein-Gordon equation becomes 
\be 
\bar{\qed}\, \phi + \bar{m}^2 c^2 \, \phi - \frac{\bar{R}}{6}\, \phi = 0,
\ee
with $\bar{\qed}$ the Laplace-Beltrami operator in the metric (\ref{confmet}). In the limit $\Lambda \to \infty$, it reduces to the conformal Klein-Gordon equation (\ref{kg2}).

\ack
The authors would like to thank FAPESP, CAPES and CNPq for financial support.


\end{document}